\documentclass[%
reprint,
superscriptaddress,
 amsmath,amssymb,
 aps,
prl,
longbibliography,
]{revtex4-1}
\usepackage{textcomp}
\usepackage{graphicx}
\usepackage{dcolumn}
\usepackage{bm}
\usepackage{color}
\usepackage{lineno}

\begin{document}

\preprint{APS/123-QED}

\title{Ultrastrongly-coupled and Directionally-nonreciprocal Magnon-polaritons in Magnetochiral Metamolecules}


\author{Kentaro Mita}
\affiliation{Department of Physics, Graduate School of Science, Tohoku University, Sendai 980-8578, Japan}

\author{Takahiro Chiba}
\affiliation{Frontier Research Institute for Interdisciplinary Sciences, Tohoku University, Sendai 980-8578, Japan}
\affiliation{Department of Applied Physics, Graduate School of Engineering, Tohoku University, Sendai 980-8579, Japan}

\author{Toshiyuki Kodama}
\affiliation{Institute for Excellence in Higher Education, Tohoku University, Sendai, 980-8576, Japan}

\author{Tetsuya Ueda}
\affiliation{Department of Electrical Engineering and Electronics, Kyoto Institute of Technology, Kyoto 606-8585, Japan}

\author{Toshihiro Nakanishi}
\affiliation{Department of Electronic Science and Engineering, Kyoto University, Kyoto 615-8510, Japan}

\author{Kei Sawada}
\affiliation{RIKEN SPring-8 Center, Sayo 679-5148, Japan}

\author{Satoshi Tomita}
\email[Email address:]{tomita@tohoku.ac.jp}
\affiliation{Department of Physics, Graduate School of Science, Tohoku University, Sendai 980-8578, Japan}
\affiliation{Institute for Excellence in Higher Education, Tohoku University, Sendai, 980-8576, Japan}

\date{\today}


\begin{abstract}
We experimentally demonstrate 
magnon-polaritons 
with ultrastrong coupling and directional nonreciprocity 
in a metamolecule 
lacking time-reversal and space-inversion symmetries 
at room temperature. 
These experimental results 
are reproduced well via numerical simulations 
and theoretical consideration. 
Ultrastrong coupling is due to a direct interaction 
of magnons in the magnetic meta-atom 
with microwave photons confined in the chiral meta-atom as a resonator. 
Our results reveal a crucial step 
in identifying deepstrongly-coupled and optically-moving magnon-polaritons 
for hybrid quantum systems, 
synthetic gauge fields, 
and quasi-particle ``chemistry'' using metamaterials.
\end{abstract}

\maketitle

\textit{Introduction} --
Condensed matter physics 
is a field of various quasi-particles with elementary excitations. 
Integrating different quasi-particles 
has led to novel and intriguing properties of matters and 
opened an essential pathway toward frontiers in physics \cite{Kockum2019,Diaz2019}. 
A magnon, a quantized elementary excitation in magnets, 
can be transformed to magnon-polariton (MP) 
when it is coupled to alternating current (AC) electromagnetic fields, 
namely photons. 
The MP has gained attention 
particularly in quantum information technology and spintronics \cite{Li2020,Harder2021}
for applications to hybrid quantum systems \cite{Lachance-Quirion2019}. 

Magnons and photons are coherently coupled 
-- strong coupling -- 
if the coupling constant, $g$ 
is larger than the dissipation rate, $\kappa$, 
in a resonator and oscillator ($g > \kappa$), 
generating MP. 
Here, 
$g$ is given by $g = (\omega_{+} - \omega_{-}) / 2$, 
where $\omega_{\pm}$ is the hybridized eigenmodes 
at the crossing point of the original modes. 
Such coherent coupling of magnons 
with photons having a frequency of $\omega / 2 \pi$ 
gives rise to the dispersion anti-crossing 
due to the hybridization. 
The anti-crossing 
is evaluated using Rabi-like level repulsion of $2g / 2\pi$, 
which is proportional to a coupling ratio of $g / \omega$. 
While the strong coupling is characterized by $g > \kappa$, 
an ultrastrong magnon-photon coupling 
is represented by $g > \kappa$ and $g / \omega > 0.1$, 
which is indispensable toward the quantum regime \cite{Bourcin2023}. 
The ultrastrongly-coupled MPs 
have been demonstrated at microwave frequencies 
using magnetic heterostructures composed of superconductors \cite{Golovchanskiy2021PRAp,Golovchanskiy2021SAd,Ghirri2023}. 
They are, however, operated at cryogenic temperatures. 
Therefore, 
operating the ultrastrongly-coupled MPs at room temperature 
remains challenging.

The ultrastrongly-coupled MP 
is inherently nonreciprocal 
owing to the breaking of time-reversal symmetry by magnetization or magnetic fields \cite{Harder2018SSP}. 
With the additional breaking of the space-inversion symmetry,
the nonreciprocity results in a transmission amplitude (and phase) difference 
depending on the propagation direction of photons. 
However, 
such a directionally-nonreciprocal optical phenomenon 
has yet to be observed in ultrastrongly-coupled systems, 
except for dissipative \cite{Harder2018PRL} 
or strong \cite{Zhang2020} magnon-photon coupling. 
The directionally-nonreciprocal phenomena exist 
in natural chiral molecules \cite{Barron1984,Rikken1997,Vallet2001}, 
multiferroic materials \cite{Okamura2015}, 
and metamaterials \cite{Tomita2014,Tomita2017,Tomita2018}
under the external direct current (DC) magnetic fields, 
as polarization-independent phenomena 
-- magnetochiral (MCh) and optical magnetoelectric (ME) effects \cite{Tokura2018,Toyoda2019}. 
Although such nonreciprocal MCh metamaterials 
indicate the presence of room-temperature coherent couplings 
between magnons and photons at microwave frequencies \cite{Tomita2017}, 
the coupling mechanism remains unclear. 
The coupling mechanism should be therefore clarified 
by evaluating the coupling ratios of the MP.

In this Letter, 
we demonstrate experimentally and numerically
ultrastrong couplings 
of the directionally-nonreciprocal MPs 
in a single three-dimensional MCh metamolecule at room temperature.
The coupling ratio is evaluated to be $g / \omega >$ 0.2, 
which is achieved 
by direct interaction 
between magnons in the yttrium-iron garnet (YIG) cylinder 
and microwave photons in the Cu chiral structure as a resonator. 
The further enhancement 
of the coupling ratio 
can be achieved by 
a decrease in the frequency of the fundamental mode of the chiral resonance, 
bringing about the deepstrong-coupling 
for hybrid quantum systems. 
Moreover, 
the metamolecule including the chiral resonator 
paves a way to the directionally-nonreciprocal MPs in the free-space 
for synthetic gauge fields for photons. 
Furthermore, 
the synthesis of metamolecules and metamaterials 
by combining different meta-atoms 
leads to quasi-particle ``chemistry''.

\begin{figure}[tb!]
\includegraphics[width=8truecm,clip]{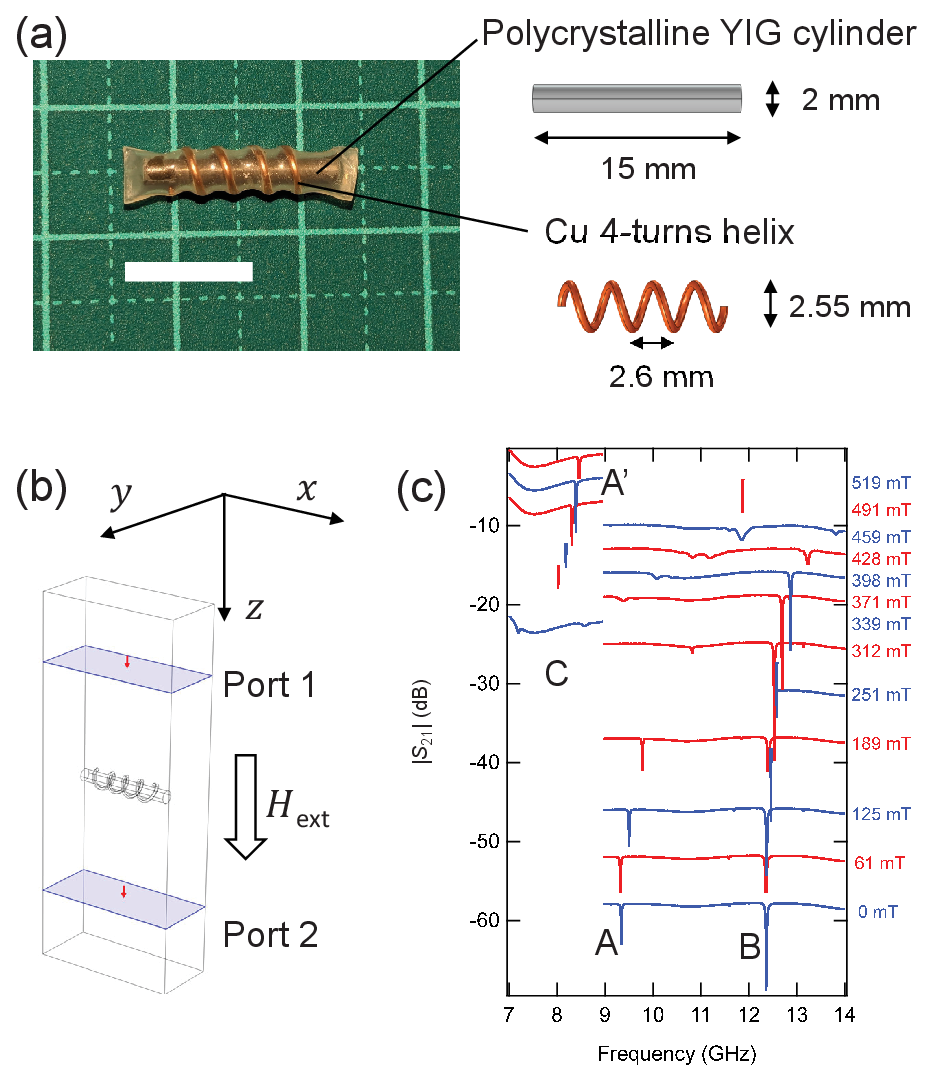}
\caption{
Magnetochiral (MCh) metamolecule under study, 
measurement setup, and measured microwave spectra of the metamolecule. 
(a) Sample photo and constituents. 
A white bar corresponding to 10 mm. 
(b) Measurement and simulation setup of the $x$-axis-oriented metamolecule placed in a waveguide. 
(c) $|S_{21}|$ amplitude spectra of the $x$-axis-oriented metamolecule at various magnetic fields.
}
\label{fig:sample}
\end{figure}

\textit{Experimental setup} --
Figure \ref{fig:sample}(a)
is an illustration of the MCh metamolecule. 
The MCh metamolecule consists of 
a polycrystalline YIG cylinder (as a magnetic meta-atom) 
inserted in a right-handed helix made of copper (Cu) (as a chiral meta-atom) \cite{Tomita2017}. 
The Cu wire of 0.55 mm diameter 
is wounded four times with a 2.6 mm pitch 
to form the right-handed helix of a 2.55 mm outer diameter. 
The YIG cylinder 
is 2 and 15 mm in diameter and length, respectively. 

As in Fig. \ref{fig:sample}(b), 
the MCh metamolecule 
is set into a WR-90 waveguide 
that supports the TE$_{10}$ mode 
with square flange adapters (Pasternack PE9804). 
The metamolecule 
is oriented along the $x$-axis, 
which is parallel to the AC magnetic field 
in the WR-90 waveguide. 
See Fig. S1 in Supplemental Material \cite{SM} 
for AC magnetic and electric fields 
of the TE$_{10}$ mode in the waveguide. 
An external DC magnetic field, $\mu_{0} H_{\rm ext}$, 
up to approximately 540 mT 
is applied in the +$z$ direction by an electromagnet. 
The waveguide is connected to 
a vector network analyzer (VNA) (Rohde \& Schwarz ZVA67) 
with a microwave input power of 0 dBm (1mW). 
The $S$ parameters 
are measured using VNA. 
The $S_{21}$ 
indicates a complex transmission coefficient from port 1 to 2, 
while $S_{12}$ represents that from port 2 to 1. 
All measurements are carried out at room temperature.

\textit{Experimental results} --
Figure \ref{fig:sample}(c) 
illustrates the transmission amplitude $|S_{21}|$ spectra 
of the $x$-axis-oriented MCh metamolecule from 7 to 14 GHz 
under varying $\mu_{0} H_{\rm ext}$ (0 - 543 mT).
Without the DC magnetic field ($\mu_{0} H_{\rm ext} = 0$), 
two sharp dips are observed: one at 9.34 GHz (A) 
and the other at 12.36 GHz (B). 
Based on numerical simulation 
using COMSOL Multiphysics of the metamolecule under $\mu_{0} H_{\rm ext} = 0$, 
we confirm that 
a dip at the lowest frequency is caused by the third-order resonance, 
with two nodes in the Cu chiral meta-atom 
induced by the microwave photons with the AC magnetic field in the $x$-axis 
(See Fig. S2(a) and Movie S1 in Supplemental Material \cite{SM}). 
The dip (A) at 9.34 GHz 
is thus traced back to the resonating photons, 
referred to as chiral resonant photons, 
because the chiral meta-atom 
functions as a resonator or sensitizer.
Contrastingly, 
the dip (B) at the higher frequency 
is not caused by simple resonance 
but by a complex mode 
with tornado-like current distribution in the Cu chiral structure 
or Mie resonance by the displaced YIG cylinder 
(Figs. S2(a) and S2(c) and Movie S2 in Supplemental Material \cite{SM}). 
Therefore, in this experiment, 
we focus on the third-order chiral resonant photon 
at approximately 9 GHz, 
which is labeled as (A). 

Applying the external DC magnetic field 
to the metamolecule in the +$z$-direction 
causes the chiral resonant photon mode (A) 
to shift slightly to a lower frequency 
at $\mu_{0} H_{\rm ext}$ = 61 mT 
and then turn to a higher frequency 
with increasing $\mu_{0} H_{\rm ext}$. 
The major blue shift 
of the chiral resonant photon mode 
is attributed to the magnetic permeability change of the YIG cylinder 
by the applied $\mu_{0} H_{\rm ext}$, 
which is the evidence of the interaction 
between the YIG magnetic meta-atom and the microwave photon in the Cu chiral meta-atom. 
When $\mu_{0} H_{\rm ext}$ = 312 mT, 
a slight dip appears 
at approximately 8 GHz labeled as (C) in Fig. \ref{fig:sample}(c). 
The slight dip keeps moving to a higher frequency 
with an increase in $\mu_{0} H_{\rm ext}$ up to 543 mT; 
therefore, 
the dip is caused by ferromagnetic resonance, 
i.e., a magnon, in the YIG magnetic meta-atom. 
When $\mu_{0} H_{\rm ext}$ = 339 mT, 
the magnon mode (C) approaches the chiral resonant photon mode (A). 
Simultaneously, 
another sharp dip appears at 7.19 GHz, 
denoted by (A'), 
and shifts to a higher frequency 
with a further increase in $\mu_{0} H_{\rm ext}$. 
This indicates 
the hybridization of chiral resonant photon modes (A) and (A') 
with the magnon mode (C), 
demonstrating the coherent coupling 
between chiral resonant photons and magnons in the metamolecule. 

\begin{figure}[tb!]
\includegraphics[width=8truecm,clip]{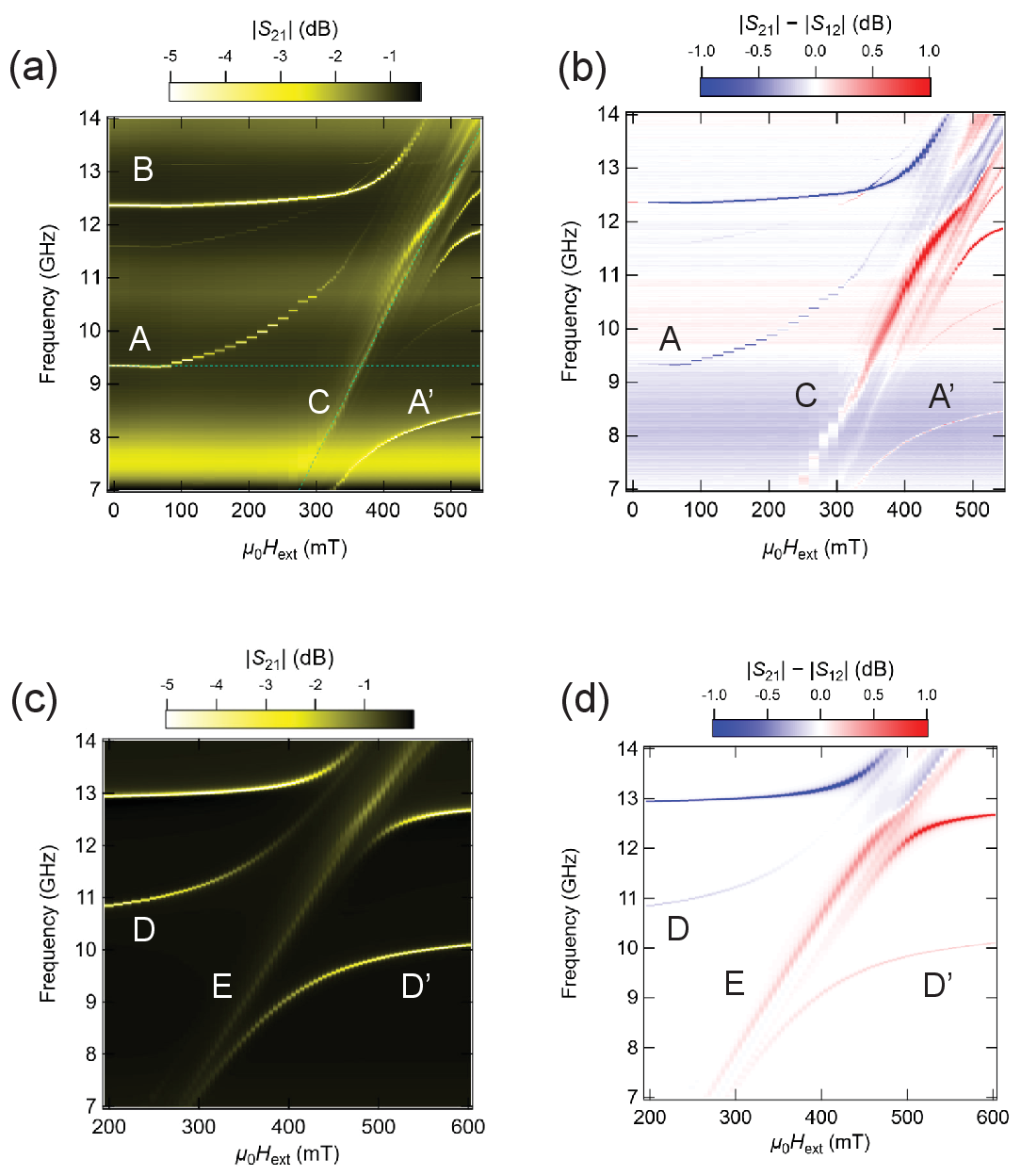}
\caption{
Two dimensional plot of measured and simulated MP's dispersions. 
Measured 2D plots of (a) transmission amplitude $|S_{21}|$ 
and (b) amplitude difference $|S_{21}| - |S_{12}|$ of $x$-axis-oriented metamolecule 
as functions of applied external DC magnetic field, $\mu_{0} H_{\rm ext}$, and frequency. 
(c) and (d) are 2D plots numerically simulated using COMSOL Multiphysics, 
corresponding to (a) and (b), respectively.
}
\label{fig:2dplot}
\end{figure}

In Fig. \ref{fig:2dplot}(a), 
$|S_{21}|$ is plotted two-dimensionally 
as functions of $\mu_{0} H_{\rm ext}$ of 0 - 543 mT (horizontal axis) 
and of frequency of 7 - 14 GHz (vertical axis) 
to evaluate magnon-photon coupling strength. 
The black color 
corresponds to a higher transmission in the $|S_{21}|$ spectra, 
while the yellow to white colors correspond to lower transmissions. 
Green dotted lines 
correspond to eye guides of 
the chiral resonant photon frequency and the magnon dispersion. 
The $|S_{21}|$ two-dimensional (2D) plot in Fig. \ref{fig:2dplot}(a) highlights, 
at $\mu_{0} H_{\rm ext} \sim$ 350 mT, 
the anti-crossing between the chiral resonant photon mode (A) 
with $\omega_{a} / 2 \pi$ = 9.34 GHz 
and magnon mode (C). 
The anti-crossing Rabi-like level splitting 
of chiral resonant photon modes (A) and (A') 
is $g / \pi$ = 4.19 GHz at $\mu_{0} H_{\rm ext}$ = 367.7 mT, 
resulting in a coupling ratio of $g / \omega_{a}$ = 0.22. 
Figure \ref{fig:2dplot}(a) demonstrates that 
magnons in the YIG cylinder and photons in the Cu chiral structure in the MCh metamolecule 
are ultra-strongly coupled at room temperature. 
The coupling ratio 
can be varied 
by the metamolecule orientation. 
Figures S3(a), S3(b), and S3(c) 
in Supplemental Material \cite{SM} 
indicate that 
the $z$-axis-orientated metamolecule 
has a smaller coupling. 
The coupling ratio 
is thus relevant to the electromagnetic mode.

Figures S4(a) and S4(b) in Supplemental Material \cite{SM} present 
the directional transmission amplitude difference, $|S_{21}| - |S_{12}|$, 
and phase difference, $\arg S_{21} - \arg S_{12} $, spectra 
of the MCh metamolecule in the $x$-axis orientation, respectively. 
Using Fig. S4(a), 
a 2D plot of experimentally observed $|S_{21}| - |S_{12}|$ 
is drawn in Fig. \ref{fig:2dplot}(b). 
The red color corresponds to $|S_{21}| > |S_{12}|$, 
while the blue color corresponds to $|S_{21}| < |S_{12}|$. 
The chiral resonant photon mode (A) above the magnon mode line (C) 
demonstrates $|S_{21}| < |S_{12}|$, 
while the mode (A') below the magnon mode (C) 
represents $|S_{21}| > |S_{12}|$. 
This is characteristic 
of the directional nonreciprocity 
of ultrastrongly-coupled MP.

\textit{Numerical simulation} --
Simultaneous breaking of time-reversal and space-inversion symmetries 
by the coexistence of magnetism and chirality in the metamolecule 
is the primary origin of the directional nonreciprocity 
observed in Fig. \ref{fig:2dplot}(b). 
An alternative and artificial origin could be 
a slight displacement of the metamolecule from the waveguide center 
or a misalignment of the metamolecule from the $x$-axis 
because similar nonreciprocal signals have been observed 
by an asymmetrical microstrip line on a ferrite substrate \cite{Ueda2006} 
and a displaced magnetic material on a coplanar waveguide \cite{Kodama2018}. 
Indeed in this study, 
numerical simulation indicates that 
the dip (B) 
observed at a higher frequency 
is caused by the Mie resonance of the displaced YIG cylinder 
(Figs. S2(a) and S2(c) in Supplemental Material \cite{SM}). 
Therefore, 
we conduct numerical simulations of the metamolecule 
under various magnetic fields 
using COMSOL Multiphysics 
to exclude such an artificial origin. 
The Supplemental Material \cite{SM} 
details the numerical simulation method 
(see also reference \cite{Pozar} therein).

Figure \ref{fig:2dplot}(c) presents 
a numerically simulated 2D plot of $|S_{21}|$ 
as functions of $\mu_{0} H_{\rm ext}$ of 200 - 600 mT (horizontal axis) 
and of 7 - 14 GHz frequency (vertical axis). 
Level repulsion due to the interaction 
between chiral resonant photons (D and D') and magnons (E) 
is reproduced. 
The simulated coupling ratio of 0.15 
derived from the level splitting of 3.21 GHz 
at $\mu_{0} H_{\rm ext}$ = 410 mT 
is smaller than that observed experimentally in Fig. \ref{fig:2dplot}(a) 
because the magnetization dynamics in the YIG cylinder 
is linearized in the numerical simulation. 
Nevertheless, 
the simulated 2D plot of $|S_{21}| - |S_{12}|$ 
in Fig. \ref{fig:2dplot}(d) 
has successfully reproduced the directional nonreciprocity 
of the MPs. 
The metamolecule 
has negligible dissipation 
and is regarded as a Hermitian system. 
As shown in Fig. \ref{fig:2dplot}(d), 
removal of the misalignment and displacement 
of the metamolecule in the simulation 
verifies that 
the origin of directionally-nonreciprocal signals 
is the MP 
under the simultaneous breaking 
of time-reversal and space-inversion symmetries in the metamolecule.

\begin{figure}[tb!]
\includegraphics[width=8truecm,clip]{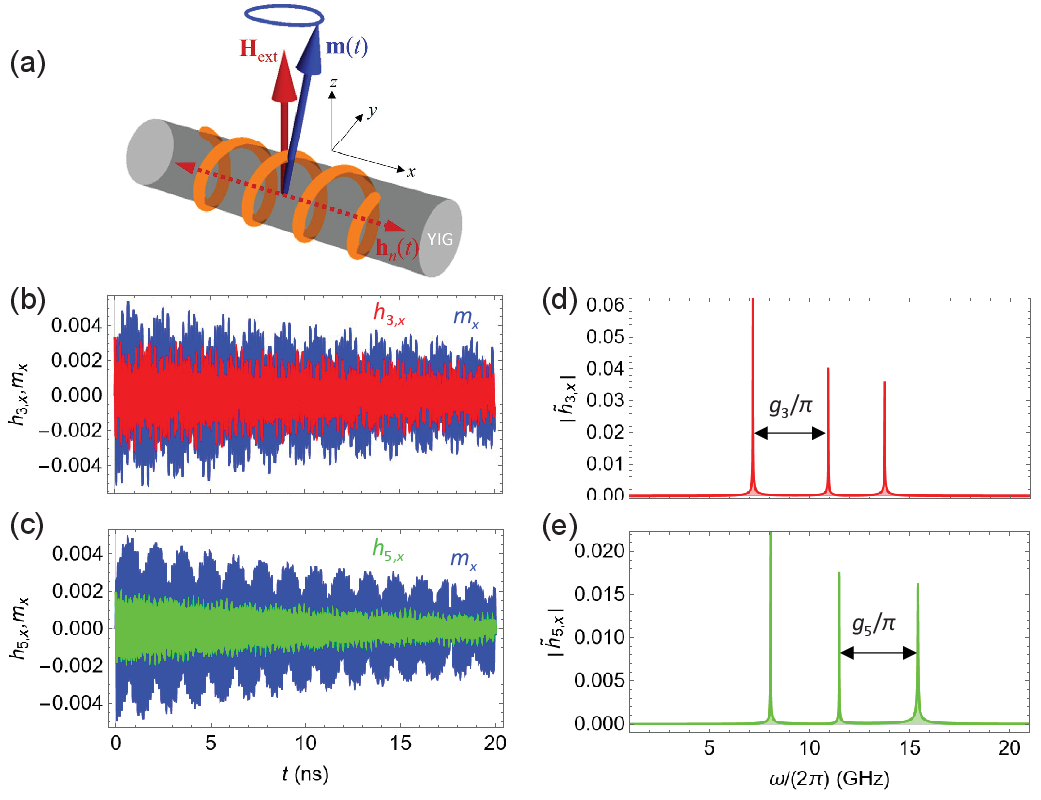}
\caption{
Theoretical consideration including magnetization dynamics. 
(a) Schematic of MCh metamolecule 
composed of a Cu chiral structure involving a YIG cylinder. 
The metamolecule's YIG cylinder is oriented to the $x$-axis. 
(b) and (c) Rabi-like oscillations of the chiral resonant photons ($h_{\rm n,x}$) 
and magnon  ($m_{\rm x}$) under $\mu_{0} H_{\rm ext}$ = 0.39 T ($n$ = 3) 
and $\mu_{0} H_{\rm ext}$ = 0.50 T ($n$ = 5), respectively. 
(d) and (e) Corresponding Fourier spectra of 
chiral resonant photons ($h_{\rm n,x}$) for 9.34 GHz and 12.36 GHz, respectively.
}
\label{fig:calc}
\end{figure}

\textit{Theoretical consideration} --
The coupling ratio of the MP 
is quantitatively reproduced 
by considering the magnetization dynamics. 
The experimental system 
is the forced oscillation of MPs 
driven by the input microwaves; 
nonetheless, a simpler damped oscillation model 
is assumed to evaluate 
the coupling ratio of the magnon and chiral resonant photons 
in the following. 
As in Fig. \ref{fig:calc}(a), 
$h_{\rm n,x}$ represents an AC magnetic field in the $x$-direction 
due to the Amp\`{e}re's circuital law 
accompanied by the Faraday's electromagnetic induction in the chiral meta-atom 
-- chiral resonant photons.
Here, the integer $n$ denotes 
the mode-index of the chiral resonant photons. 
The AC magnetic field $h_{\rm n,x}$ 
excites magnons ($m_{\rm x}$) in the YIG cylinder. 
This leads to direct coupling 
between magnons and chiral resonant photons, 
generating ultrastrongly-coupled MPs in the metamolecule. 
Recall that the numerical simulation using COMSOL 
indicates that the third-order resonance with $n$ = 3 ($h_{\rm 3, x}$) 
appears at approximately 10 GHz.  
Thus, 
the MPs 
can be described as a coupled oscillator model 
that considers a uniform magnon mode and two chiral resonant modes with $n$ = 3 and 5. 
The Supplemental Material \cite{SM}
presents the calculation details 
(see also reference \cite{Cao2015} therein).

Figures \ref{fig:calc}(b) and \ref{fig:calc}(c) 
are calculated damped oscillations 
of the chiral resonant photons ($h_{\rm n, x}$) 
and the magnons ($m_{x}$) 
under $\mu_{0} H_{\rm ext}$ = 0.39 T for $n$ = 3 
and under $\mu_{0} H_{\rm ext}$ = 0.50 T for $n$ = 5, respectively. 
We assume 
the higher frequency mode to be the fifth-order resonance. 
The time evolutions in Figs. \ref{fig:calc}(b) and \ref{fig:calc}(c) 
correspond to Rabi-like oscillations 
-- the amplitude exchange between one chiral resonant photon (red or green) 
and a magnon (blue). 
A minimal complexity in the Rabi-like oscillations 
is attributed to interactions among three modes 
(one magnon and two chiral resonant photons) 
in the metamolecule. 

Figures \ref{fig:calc}(d) and \ref{fig:calc}(e) 
correspond to the Fourier spectra of the time evolutions, 
in which three peaks represent coupled modes 
among one magnon and two chiral resonant photons, 
i.e., MP modes with the Rabi-like frequency splitting. 
According to these spectra, 
the evaluated couplings are 
$g_{3} / \pi$= 3.75 GHz and $g_{5} / \pi$ = 3.93 GHz, 
resulting in the coupling ratios 
of $g_{3} / \omega_{3}$ = 0.20 and $g_{5} / \omega_{5}$= 0.16, 
respectively. 
Here, 
the subscripts in $g_{n}$ and $\omega_{n}$ 
mean the corresponding values of the $n$-th order. 
Based on the experimental results, 
$\omega_{3} / 2 \pi$ = 9.34 and $\omega_{5} / 2 \pi$ = 12.36 GHz 
are assumed for this evaluation. 
The computed $g_{3} / \omega_{3}$ = 0.20 
is consistent with the experimentally obtained $g / \omega_{\rm a}$ = 0.22, 
indicating that 
MP in the MCh metamolecule 
is in the ultrastrong coupling regime. 
Moreover,  
$g_{3} / \omega_{3}$ = 0.20 and  $g_{5} / \omega_{5}$ = 0.16 
indicate that 
the coupling at a lower frequency is much stronger than that at a higher frequency.

\textit{Discussion} --
This study reveals that 
the MCh metamolecule 
consisting of a YIG cylinder inserted in a Cu chiral structure 
results in ultrastrongly-coupled and directionally-nonreciprocal MP 
at room temperature. 
Table S1 in Supplemental Material \cite{SM}
compares the temperature, MP coupling ratio, and nonreciprocity 
obtained in this study with those reported in the literature 
(see also references \cite{Zhang2014,Bai2015} therein).
Ultrastrong coupling in the MP 
is primarily caused by direct interaction 
between photons in the Cu chiral structure and 
magnons in the magnetic YIG cylinder 
via AC magnetic fields in the $x$-direction.
The coupling ratio 
is reduced in the $z$-axis-orientated metamolecule (Fig. S3(c)) 
because electromagnetically induced AC magnetic fields in the $z$-direction 
cannot drive magnons 
under the external DC magnetic field, $\mu_{0} H_{\rm ext}$, in the $z$-direction. 
This results in 
indirect coupling between the chiral resonant photon and magnon 
in the $z$-axis-orientated metamolecule. 
Therefore, 
the coupling ratio can be enhanced 
by finding a suitable metamolecule orientation.

Utilizing the fundamental mode ($n$ = 1) 
is another strategy toward achieving a much stronger coupling 
between magnons and photons 
–- the deepstrongly-coupled MP, 
i.e., $g / \omega_{\rm a} \geq$ 1, 
in the metamolecule. 
This experiment 
uses the third-order mode ($n$ = 3) of the chiral resonant photon 
to obtain the highest value of $g / \omega_{\rm a}$ = 0.22; 
the magnon-photon coupling is thus reduced by $\eta_{3}$ = 1/3, 
where $\eta_{\rm n}$ is a phenomenologically introduced mode-dependent coefficient 
responsible for the space-dependence 
of the microwave magnetic field in the MCh metamolecule. 
Here, 
we assume the net magnitude 
of the AC magnetic field of the chiral resonant photon 
to be inversely proportional to the number of antinodes 
(See more detailed in Supplemental Material \cite{SM}). 
In this way, 
the $n$ = 1 fundamental mode 
may enhance the magnon-photon coupling 
by maximizing the electromagnetic induction due to the magnon dynamics.

Additionally, 
according to the theoretical prediction for the metamolecule \cite{Chiba2024}, 
the coupling ratio is evaluated via $g / \omega_{\rm a} \propto 1/\sqrt{\omega_{\rm a}}$. 
Therefore, 
further enhancement toward the deepstrong-coupling 
is possible by decreasing the chiral resonant frequency, $\omega_{\rm a}$, 
of the fundamental mode 
by increasing the system size 
and changing the design of the metamolecule 
and materials of the constitutive meta-atoms. 
The deepstrongly-coupled MP 
will lead to mysterious quantum effects, 
such as the vacuum Bloch-Siegert shift \cite{Li2018} 
and quantum squeezing under thermal equilibrium \cite{Hayashida2023}. 
The Hermitian, 
ultrastrongly-coupled, 
and directionally-nonreciprocal MP 
may give rise to 
coherent quantum rectification of quantum information.

The second origin 
of the ultrastrong coupling in this study 
is the photon confinement in the chiral meta-atom. 
The chiral meta-atom 
functions as a resonator or sensitizer; 
because the resonator is embedded in the metamolecule 
that supports MPs, 
the MPs for propagating microwaves in the free-space 
can be realized using the metamolecule. 
As shown in Fig. S5 in Supplemental Material \cite{SM}, 
our additional numerical simulation using COMSOL 
verifies the free-space MP 
with ultrastrong coupling and directional nonreciprocity. 
The MP in the free-space 
is a central topic in quantum optics 
as it allows control over individual quantum systems \cite{Bayer2017,Rajabali2022}. 
Moreover, 
the nonreciprocal MP in the free-space 
leads to the observation of the polarization-independent optically-moving effects. 
The optically-moving effect in the MCh and ME media 
exhibits an electromagnetic response 
equivalent to that of a regular material moving at a relativistic speed \cite{Kong,Asadchy2018,kodama2024}. 
Therefore, 
free-space MPs with ultrastrong coupling and directional nonreciprocity
enable us to realize synthetic gauge fields, 
i.e., the Lorentz force acting on the propagating microwaves \cite{Sawada05}.

Last but not least, 
the quasi-particle integration 
is comprehensively classified 
according to the interaction 
between polarization $P$ and magnetization $M$. 
The conventional ME and MCh effects 
observed in this study 
are given by the coupling between $P$ and $M$. 
Contrastingly, 
the dynamic coupling of time-derivatives of $P$ and $M$ 
results in a new collective mode 
corresponding to an MCh electromagnon-polariton, 
which is still unclear but anticipated to be observed using metamaterials. 
Furthermore, 
we focus on magnons in this study; 
however, 
other quasi-particles, 
such as, excitons, phonons, and plasmons, 
can be integrated in the metamaterials. 
Thus, 
this study signifies a pivotal advancement 
toward expounding quasi-particle ``chemistry'' using metamaterials.

\textit{Conclusion} --
We demonstrate 
ultrastrong couplings 
of the directionally-nonreciprocal MPs 
in a single three-dimensional MCh metamolecule at room temperature.
The metamolecule 
exhibits the coupling ratio of $g / \omega$ = 0.22, 
which is achieved by direct interaction 
between magnons in the YIG meta-atom and 
microwave photons confined in the Cu chiral meta-atom as the resonator. 
These experimental results 
are reproduced well via numerical simulations and theoretical consideration.
The coupling ratio and directional nonreciprocity 
can be enhanced by changing the metamolecule orientation and 
the chiral resonance frequency.
The MCh metamolecule 
represents a crucial step toward developing the hybrid quantum systems 
and quasi-particle ``chemistry''using metamaterials.
Furthermore, 
the optically-moving MPs by the metamolecules in the free-space 
enable us to realize the Lorentz force acting on the propagating microwaves.

\textit{Acknowledgements} --
We thank 
H. Yasuda and Y. Motoda for their technical support 
with the microwave measurements 
and H. Kurosawa for helping 
with the numerical simulation.
This work is financially supported 
by JSPS-KAKENHI (JP24H02232, 23K13621, 22K14591) 
and JST- CREST (JPMJCR2102).



\end{document}